\documentclass[aps,superscriptaddress,showpacs,twocolumn,floatfix]{revtex4}

\usepackage{amssymb}
\usepackage{amsmath}
\usepackage{graphicx}
\usepackage[normalem]{ulem}
\usepackage[dvips]{color}
\usepackage{multirow}
\usepackage{dcolumn}                 
\usepackage[bookmarksnumbered,bookmarksopen,colorlinks,citecolor=blue,linkcolor=blue]{hyperref} %
\usepackage{CJK}                     
\usepackage[normalem]{ulem} 
\usepackage[dvips]{color} 
\renewcommand\sout{\bgroup \color{red} \ULdepth=-.5ex \ULset}

\usepackage{epstopdf, color}

\def\rpi {$\pi^-/\pi^+$~}
\def\es0{$E_{sym}(\rho_0)$~}
\begin{document}
\begin{CJK*}{GBK}{song}
\title{Effects of an induced electric field on \rpi ratio in heavy-ion collisions}
\author{Gao-Feng Wei}\email[Corresponding author. E-mail: ]{wei.gaofeng@foxmail.com}
\affiliation{Shaanxi Key Laboratory of Surface Engineering and Remanufacturing, School of Mechanical and Material Engineering, Xi'an University, Xi'an 710065, China}
\affiliation{State Key Laboratory of Theoretical Physics, Institute of Theoretical Physics, Chinese Academy of Sciences, Beijing 100190, China}
\author{Shi-Hai Dong}\email[E-mail: ]{dongsh2@yahoo.com}
\affiliation{CIDETEC, Instituto Polit\'ecnico Nacional, Unidad Profesional ALM, Mexico D. F. 07700, Mexico}
\author{Xin-Wei Cao}
\affiliation{School of Mechanical and Material Engineering, Xi'an University, Xi'an 710065, China}
\author{Yun-Liang Zhang}
\affiliation{School of Information Engineering, Xi'an University, Xi'an 710065, China}


\begin{abstract}
Using an isospin- and momentum-dependent transport model, we examine the effects of an electric field
induced by a variable magnetic field on the \rpi ratio in central to peripheral heavy-ion collisions
at beam energies of 400 and 1500MeV/nucleon. It is shown that while the induced electric
field does not affect the total multiplicities of both $\pi^{-}$ and $\pi^{+}$ mesons at both the
lower beam energy of 400MeV/nucleon and the higher beam energy of 1500MeV/nucleon, it reduces (enhances)
the emission of $\pi^{-}$ ($\pi^{+}$) mesons in midrapidity, but enhances (reduces) the emission of $\pi^{-}$
($\pi^{+}$) mesons in forward and backward rapidities especially for the more peripheral collisions
at the lower beam energy because of the rapidly transient variable magnetic field at more peripheral
collisions and longer reaction duration time at the lower beam energy. These findings indicate that
the total \rpi ratio is still a precisely reliable probe of symmetry energy at both the lower and higher
beam energies, but one should consider the induced electric field when using the differential \rpi ratio
to probe the symmetry energy especially for the lower beam energy and more peripheral collisions.
Finally, the relative suppression factor based on the ratio of \rpi in different rapidities is proposed
to be an effective probe of the induced electric field generated in heavy-ion collisions due to its
maximizing effects of induced electric fields on the differential \rpi ratio but minimizing effects of
some uncertainty factors in heavy-ion collisions.

\end{abstract}

\maketitle

\section{Introduction}\label{introduction}
Heavy-ion collision induced by a radioactive beam is one of the effective methods to determine the
density dependence of nuclear symmetry energy in experiment and theory. However, the determination
of nuclear symmetry energy requires one to compare continuously theoretical simulations with experimental
measurements. Up to now, while efforts such as the FOPI \cite{FOPI} and FOPI-LAND
\cite{FOPI-LAND1,FOPI-LAND2} experiments have provided some data involving the probe of symmetry energy
such as the \rpi ratio and the elliptic flow of nucleon, the corresponding theoretical simulations do not give
consistent conclusions on the determination of symmetry energy especially for the determination of
high-density symmetry energy. Without considering the possibly introduced errors due to
experimental environment, the main reasons are the possible uncertainty factors in theoretical predictions
and/or model dependence in different theoretical communities. For example, by comparing the \rpi ratio with
the FOPI experimental data, the Boltzmann-Uehling-Uhlenbeck (BUU) \cite{Xiao09} and Boltzmann-Langevin
(BL) \cite{Xie13} communities favor a supersoft symmetry energy, but the quantum molecular dynamics (QMD)
\cite{Feng10} community suggests a superstiff symmetry energy. Therefore, one needs to check these possible
uncertainty factors of theoretical predictions to better determine the density dependence of symmetry energy.
Presently, many communities have attempted to solve this problem including studying the impact of in-medium
pion potential \cite{Hong14,Guo15a}, the isovector potential of $\Delta$(1232) resonance \cite{Bao15a,Guo15b},
medium modification of pion production threshold \cite{Song15}, potential modification of local
and/or global energy conservation of binary collisions \cite{Cozma16}, and tensor-force-induced
short-range correlations \cite{Orhen15,Bao15b}.

On the other hand, electromagnetic interactions play an important role in the time evolution of charged particles
in heavy-ion collisions. For example, the Coulomb field of spectators in peripheral collisions can lead
to the well-known Coulomb peaks in the rapidity distribution of \rpi ratio \cite{Wei14},
and the elliptic flow splitting of charged pion mesons measured by STAR Collaboration
\cite{Wang13,Ke12,Adam15,Huang15} is closely related to chiral magnetic effects in relativistic energy
collisions. Moreover, it has been reported that a strong magnetic can be generated in the direction
perpendicular to the reaction plane (i.e., $y$ direction), regardless of intermediate energy or relativistic
energy collisions \cite{Ou11,Deng12}. Nevertheless, as has been guessed in Ref. \cite{Liu12},
the violent variable magnetic field in $y$ direction may generate a strong electric field, i.e., an induced
electric field, in relativistic heavy-ion collisions. Certainly, the changing of magnetic field in
intermediate energy collisions is less violent than that in relativistic energy collisions; thus the
effects of the induced electric field on heavier charged particle are expected not to be obvious. However,
due to the longer duration time of the induced electric field acting on charged particles compared
to the case of relativistic energy collisions, it is therefore expected that the induced electric field
will affect the lighter pion mesons and their \rpi ratio to some extent.
In addition, while almost all the transport models have consistently concluded that the \rpi ratio is
one of the most promising probes of symmetry energy especially for the high-density behavior of
symmetry energy \cite{Bar05,Bao08}, the comparison of theoretical predictions with exiting data is still
inconclusive. Therefore, to extract reliably precise information from the \rpi ratio about the high-density
symmetry energy, one needs to answer precisely whether and how the induced electric field affects
the charged pion mesons and their \rpi ratio.
To this end, we perform the Au+Au collisions to show the effects of the induced electric field on charged
pion mesons and their \rpi ratio in intermediate energy heavy-ion collisions. It is shown later
that while the induced electric field does not affect the total \rpi ratio, it suppresses
(enhances) differential \rpi ratio in midrapidity (forward and backward rapidities) especially
for very peripheral heavy-ion collisions. Therefore, it can be concluded that the total \rpi ratio
as the probe of symmetry energy is more delicate than the differential \rpi ratio due to the
considered induced electric field factor.


\section{Induced electric field in IBUU model}\label{model}
The present study is based on an isospin-dependent Boltzmann-Uehling-Uhlenbeck (IBUU) transport
model \cite{IBUU}. In this model, an isospin-dependent mean-field is used to model the nuclear
interaction; its expression is defined as follows:
\begin{eqnarray}
U(\rho,\delta ,\vec{p},\tau ) &=&A_{u}(x)\frac{\rho _{-\tau }}{\rho _{0}}%
+A_{l}(x)\frac{\rho _{\tau }}{\rho _{0}}  \notag \\
&+&B(\frac{\rho }{\rho _{0}})^{\sigma }(1-x\delta ^{2})-8\tau x\frac{B}{%
\sigma +1}\frac{\rho ^{\sigma -1}}{\rho _{0}^{\sigma }}\delta \rho
_{-\tau }
\notag \\
&+&\frac{2C_{\tau ,\tau }}{\rho _{0}}\int d^{3}p^{\prime }\frac{f_{\tau }(%
\vec{p}^{\prime })}{1+(\vec{p}-\vec{p}^{\prime })^{2}/\Lambda ^{2}}
\notag \\
&+&\frac{2C_{\tau ,-\tau }}{\rho _{0}}\int d^{3}p^{\prime }\frac{f_{-\tau }(%
\vec{p}^{\prime })}{1+(\vec{p}-\vec{p}^{\prime })^{2}/\Lambda ^{2}}.
\label{MDIU}
\end{eqnarray}%
In the above, $\rho=\rho_n+\rho_p$ is the nucleon number density and $\delta=(\rho_n-\rho_p)/\rho$
is the isospin asymmetry of the nuclear medium; $\rho_{n(p)}$ denotes the neutron (proton) density,
the isospin $\tau$ is $1/2$ for neutrons and $-1/2$ for protons, and $f(\vec{p})$ is the local phase
space distribution function. The expressions and values of the parameters $A_{u}(x)$, $A_{l}(x)$,
$\sigma$, $B$, $C_{\tau ,\tau }$, $C_{\tau ,-\tau }$, and $\Lambda $ can be found in
Refs. \cite{Das03,Che05}, and they lead to the binding energy of $-16$ MeV, incompressibility
$212$ MeV for symmetric nuclear matter, and the symmetry energy $E_{sym}(\rho_0)=30.5$ MeV at saturation
density $\rho_0=0.16$ fm$^{-3}$. The parameter $x$ is used to mimic the different
forms of symmetry energy predicted by various many-body theories without changing any property
of symmetric nuclear matter and the value of symmetry energy at saturation density.

\begin{figure}[b]
\centerline{\includegraphics[width=1.1\columnwidth]{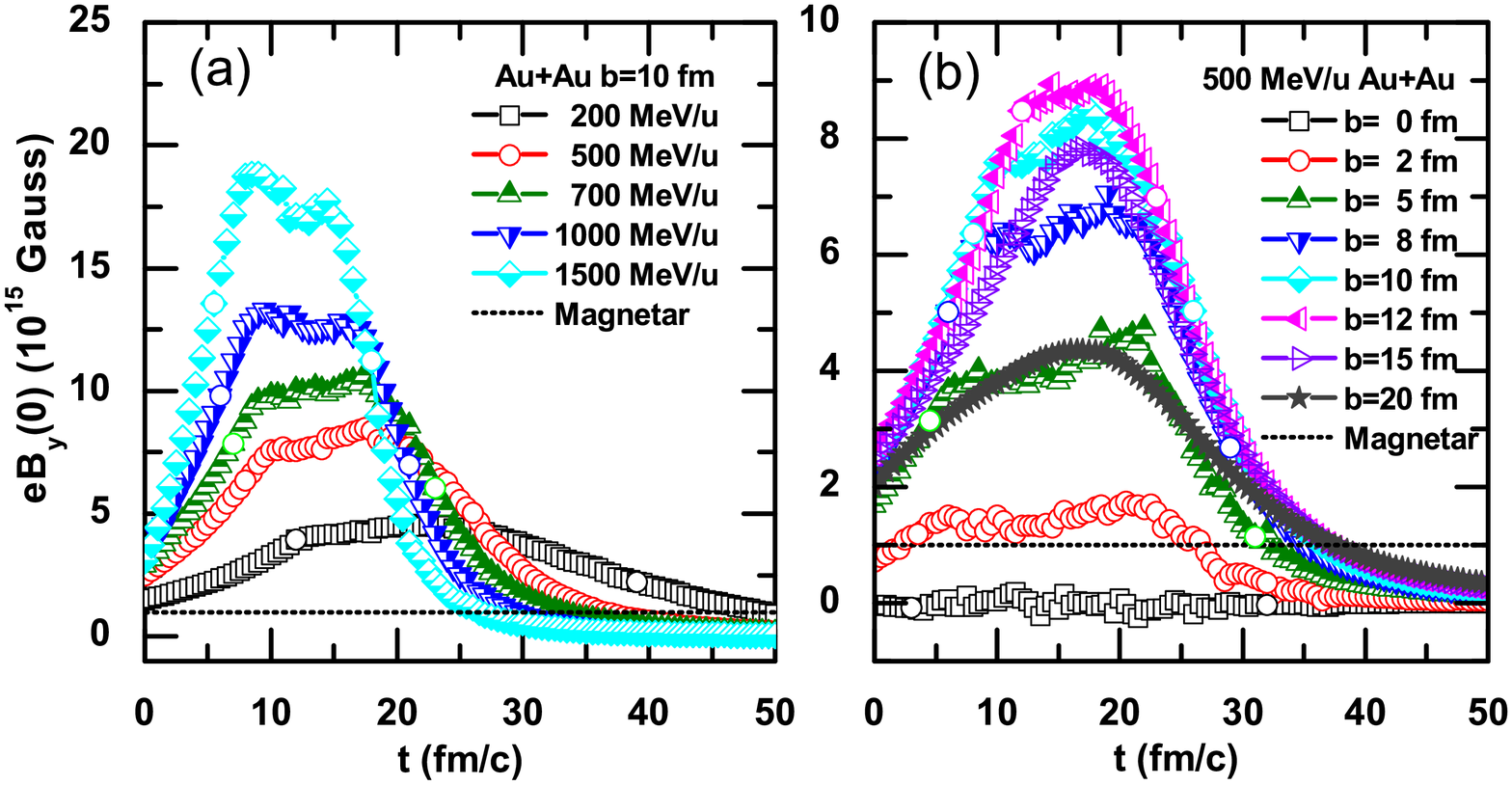}}
\caption{(Color online) The beam energy dependence of the central magnetic field $eB_{y}(0)$ in
Au+Au collisions at the impact parameter of 10fm (a), and the impact parameter dependence of
the central magnetic field $eB_{y}(0)$ in Au+Au collisions with a beam energy of 500MeV/nucleon (b).
Data are taken from Ref. \cite{Ou11}. } \label{em}
\end{figure}

\begin{figure}[h]
\centerline{\includegraphics[width=1.0\columnwidth]{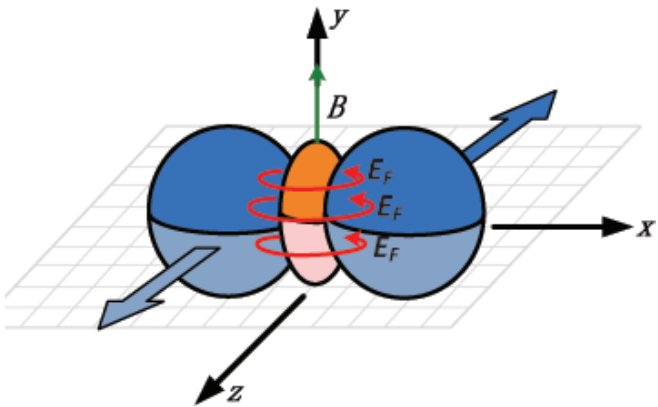}}
\caption{(Color online) A sketch of the induced electric field $\vec{E}_{F}$ due to variable magnetic
field $\vec{B}$ in the direction perpendicular to the reaction plane.
Modified from Fig. 1 of Ref. \cite{Liu12}. } \label{ef}
\end{figure}
In the IBUU model, the Li\'{e}nard-Wiechert electromagnetic potentials at the position $\vec{r}$ and time $t$
are included consistently to satisfy Maxwell's equations \cite{Ou11,Deng12},

\begin{equation}
e\vec{E}(\vec{r},t)=\frac{e^2}{4\pi \varepsilon_{0}}
\sum_{n}Z_{n}\frac{c^2-v^{2}_{n}}{(cR_{n}-\vec{R}_{n}\cdot \vec{v}_{n})^3}(c\vec{R}_{n}-R_{n}\vec{v}_{n}),
\end{equation}
\begin{equation}\label{mforce}
e\vec{B}(\vec{r},t)=\frac{e^2}{4\pi \varepsilon_{0}}
\sum_{n}Z_{n}\frac{c^2-v^{2}_{n}}{(cR_{n}-\vec{R}_{n}\cdot \vec{v}_{n})^3}\vec{v}_{n}\times \vec{R}_{n},
\end{equation}
where $Z_{n}$ is the charge number of the $n$th particle at the location of $\vec{r}_{n}$, and
$\vec{R}_{n}=\vec{r}-\vec{r}_{n}$ is the relative position of the field point $\vec{r}$ to the source
point $\vec{r}_{n}$. The summation runs over all charged particle with velocity of $\vec{v}_{n}$ at
the retarded time of $t_{n}=t-|\vec{r}-\vec{r}_{n}|$. As has been reported in Ref. \cite{Ou11,Deng12},
the dominant component of the magnetic fields is in the direction perpendicular to reaction plane, i.e.,
$y$ direction, and the component of the magnetic field in reaction plane due to event-by-event fluctuation of
the positions of charged particles is negligible due to the random motions of nucleons in $x$ or $y$
direction. Shown in Fig. \ref{em} is the beam-energy dependence of the central magnetic field in
Au+Au collisions at the impact parameter of 10 fm [panel (a)], and the impact-parameter dependence of
the central magnetic field in Au+Au collisions with a beam energy of 500MeV/nucleon [panel (b)]. It can
be seen that the magnetic field increases with increasing impact parameter and becomes more
and more steep with increasing beam energy. This rapidly changing magnetic field naturally generates a
transient electric field circling the $y$ axis in the $x$-$z$ plane as shown in Fig. \ref{ef}.
According to Faraday's law, the magnitude of the induced electric field can be evaluated by
\begin{equation}\label{ief}
|e\vec{E}_{F}|=\frac{1}{2\pi r}\frac{\Delta \phi}{\Delta t},
\end{equation}
where $r$ is the distance from the $y$ axis, and $\Delta \phi$ denotes the change of magnetic
flux through area $\pi r^2$ in the time interval $\Delta t$. Considering that the changing of
the magnetic field in intermediate energy collisions is less violent than that in relativistic-energy 
collisions, thus the effects of the induced electric field on heavier charged particles
are unlikely to be obvious. Nevertheless, due to the longer duration time of the induced
electric field acting on charged particles compared to the case of relativistic-energy collisions,
it is thus expected that the induced electric field may affect the lighter pion mesons and
their \rpi ratio to some extent. For this purpose, we simply fix the parameter $x$ at the value
of 1 (i.e., the value of 16.4MeV for the density slope of symmetry energy at normal density) to
see whether the pion mesons and their \rpi ratio do change or not under the consideration of the induced
electric field.
\section{Results and Discussion}\label{results}
\begin{figure}[b]
\centerline{\includegraphics[width=1.2\columnwidth]{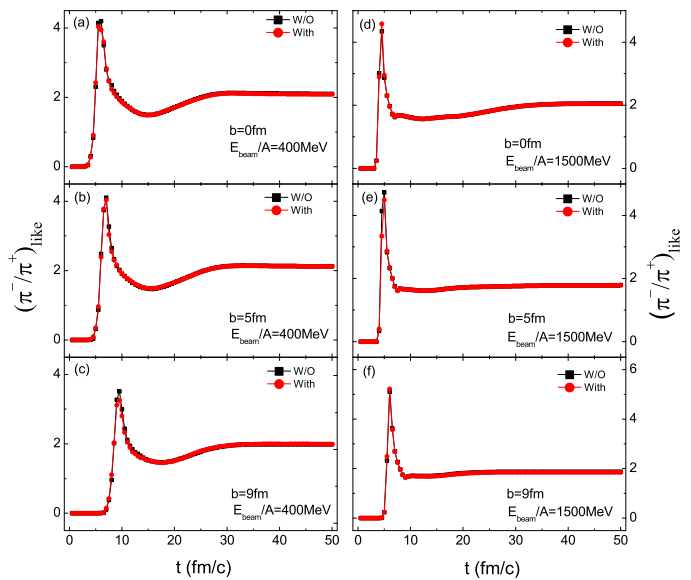}}
\caption{(Color online) The time evolution of impact-parameter-dependent dynamic (\rpi)$_{like}$ ratio
in Au+Au collisions with (labeled as With) and without (labeled as W/O) the consideration of
the induced electric field at two beam energies of
400 MeV/nucleon (left panel) and 1500 MeV/nucleon (right panel). } \label{tpion}
\end{figure}

\begin{figure}
\centerline{\includegraphics[width=1.0\columnwidth]{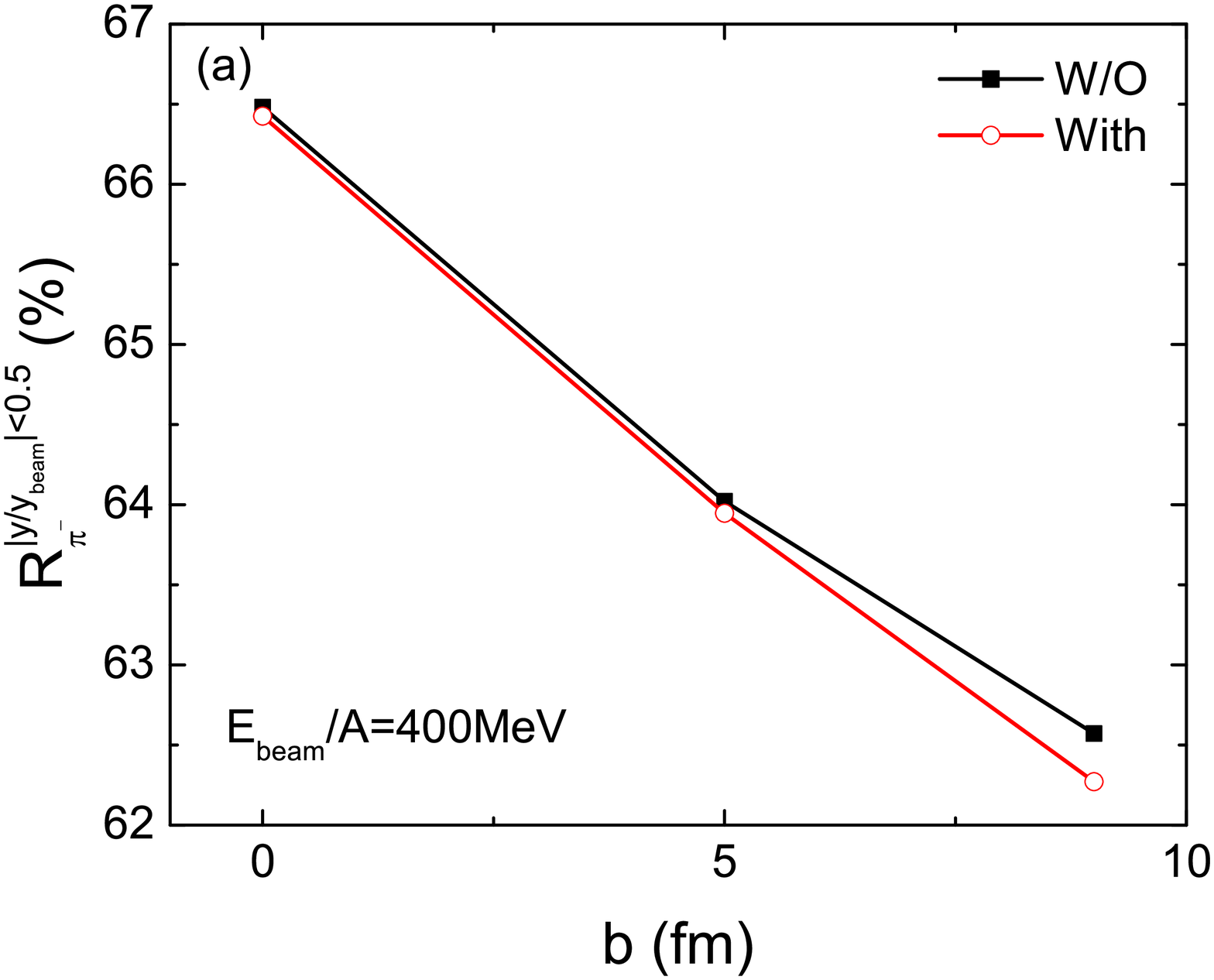}}
\centerline{\includegraphics[width=1.0\columnwidth]{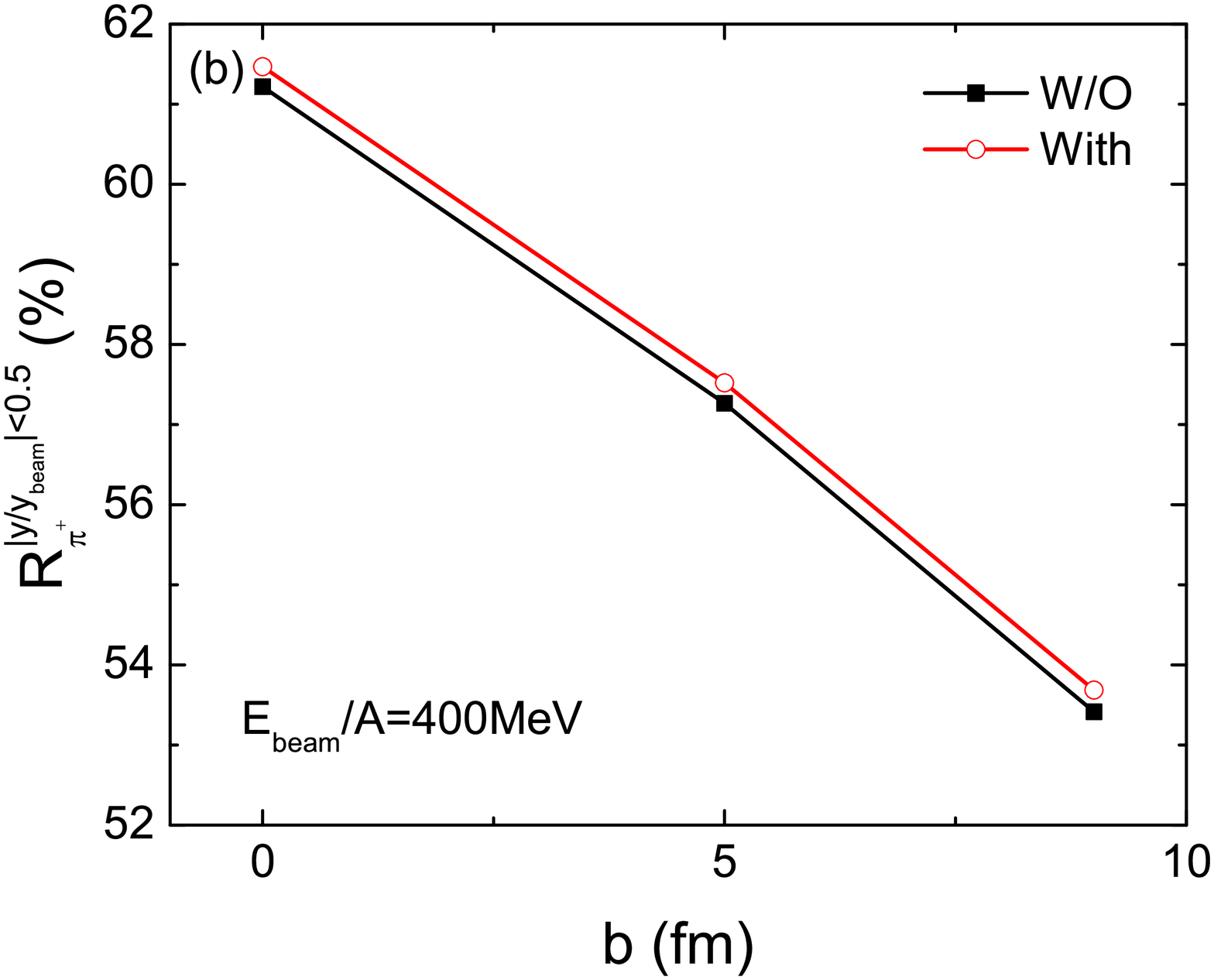}}
 \caption{(Color online) The impact-parameter-dependent percentage of $\pi^{-}$ (a) and $\pi^{+}$ (b) multiplicities
of midrapidity in Au+Au collisions with (labeled as With) and without (labeled as W/O) the consideration of
the induced electric field at a beam energy of 400 MeV/nucleon. } \label{pimid1}
\end{figure}
\begin{figure}
\centerline{\includegraphics[width=1.0\columnwidth]{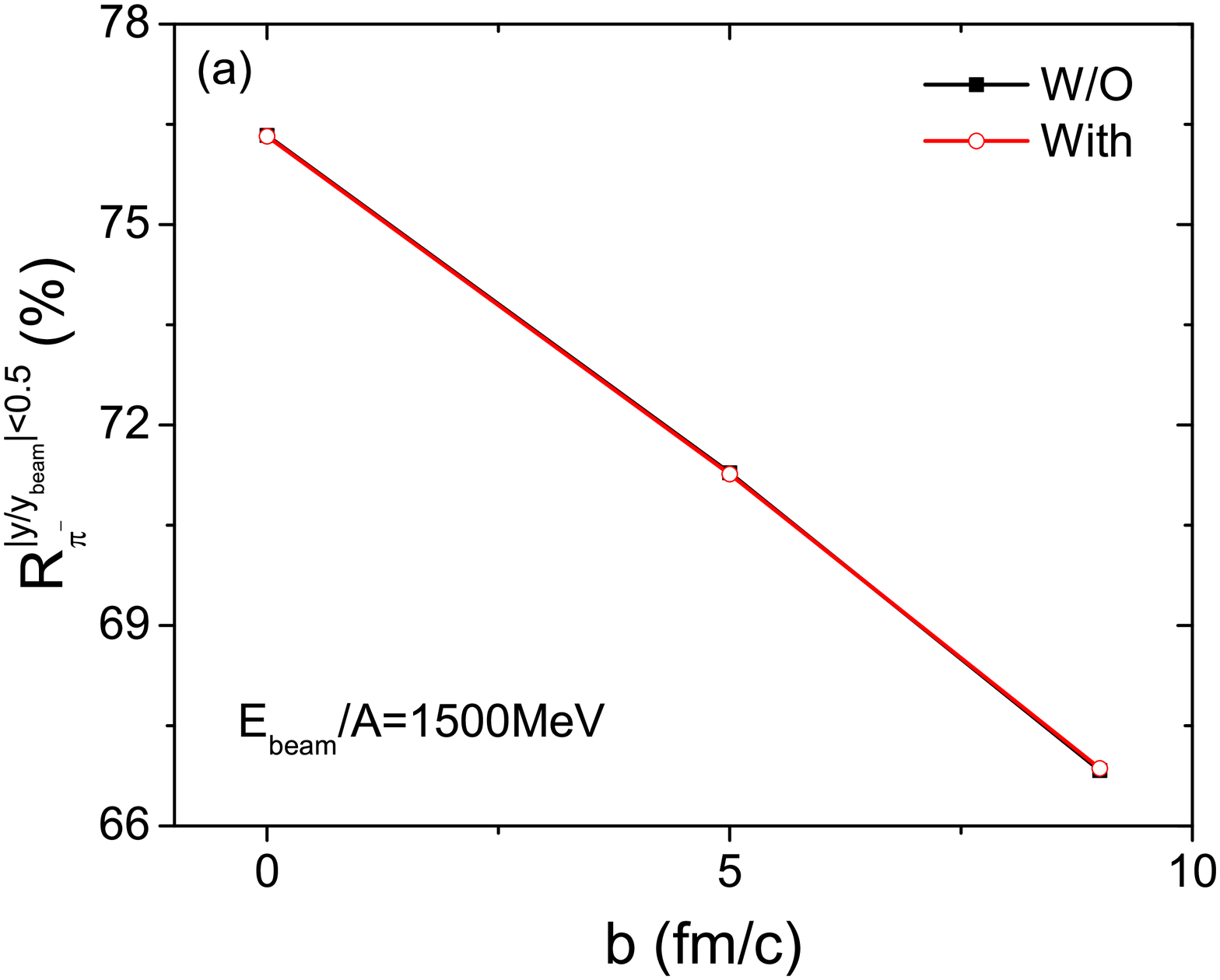}}
\centerline{\includegraphics[width=1.0\columnwidth]{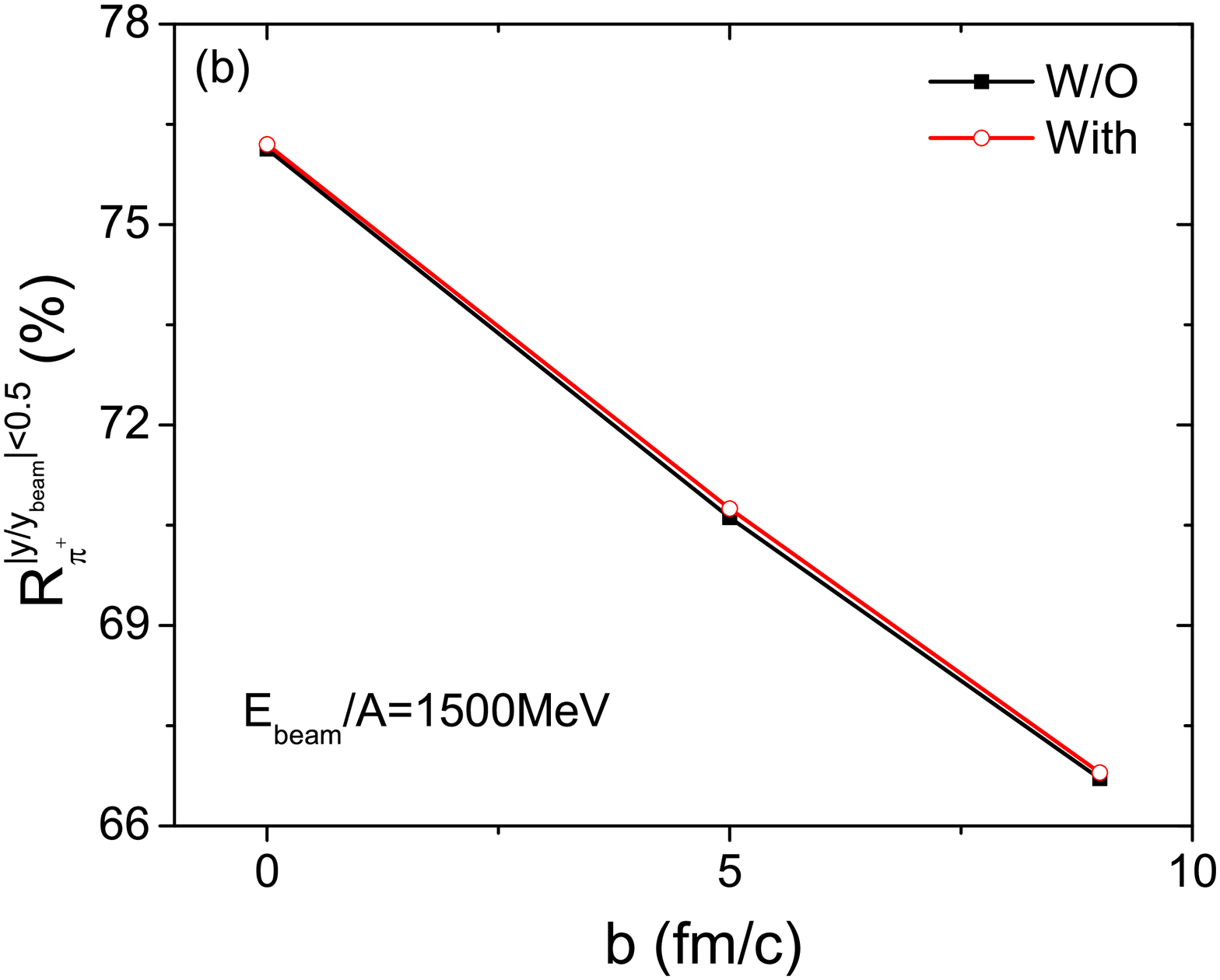}}
 \caption{(Color online) Same as Fig. \ref{pimid1} but for a beam energy of 1500MeV/nucleon. } \label{pimid2}
\end{figure}
\begin{figure}
\centerline{\includegraphics[width=1.0\columnwidth]{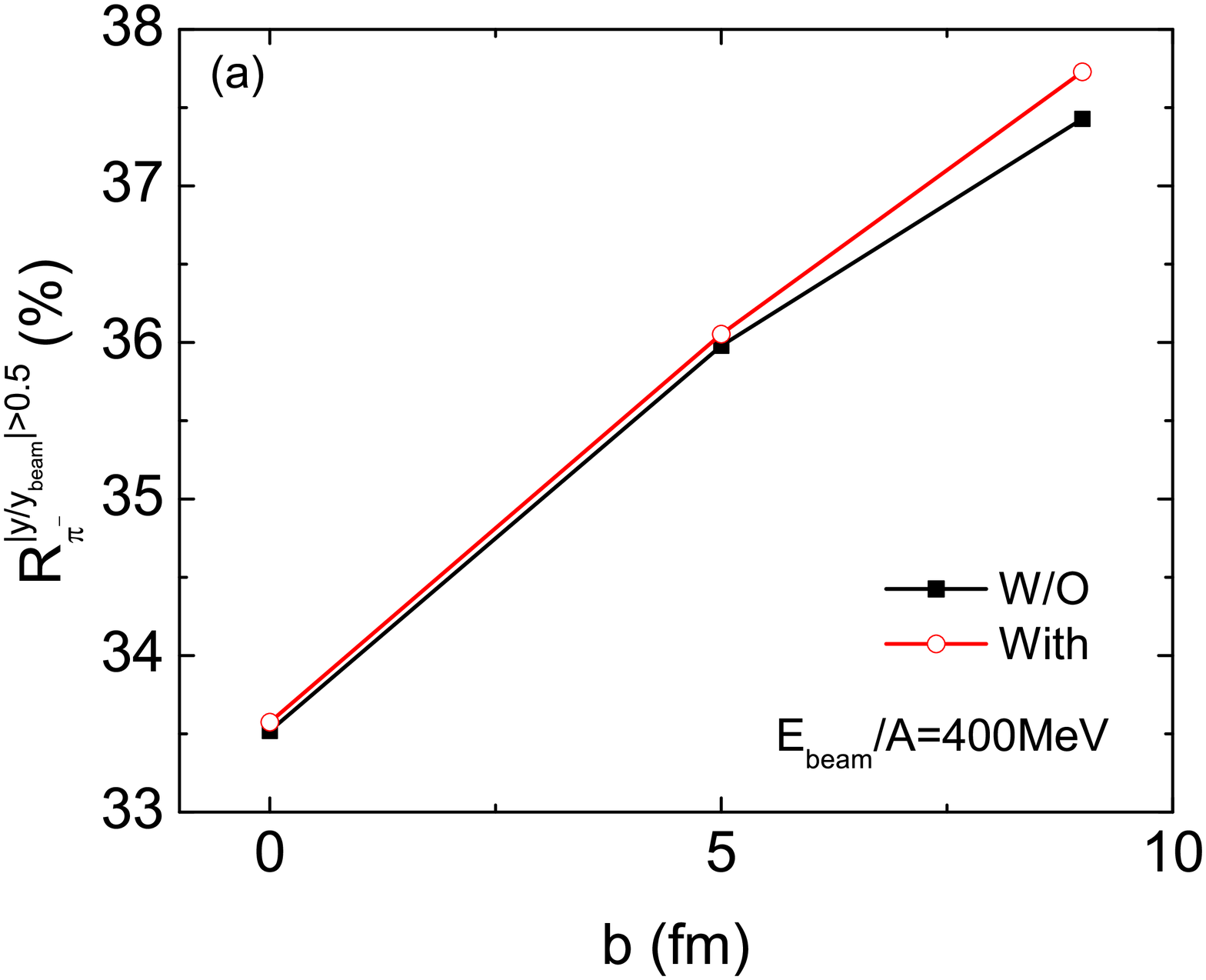}}
\centerline{\includegraphics[width=1.0\columnwidth]{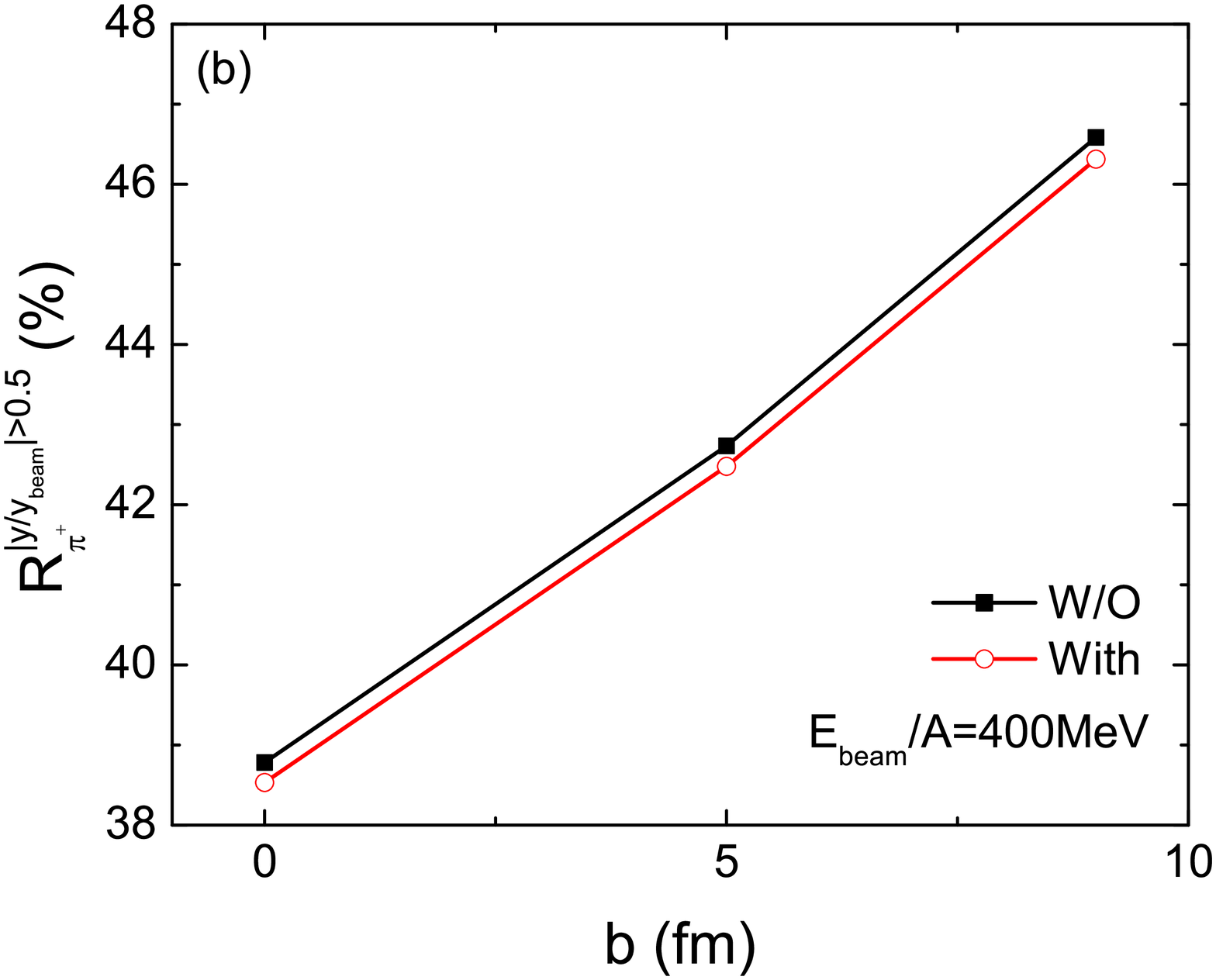}}
 \caption{(Color online) The impact-parameter-dependent percentage of $\pi^{-}$ (a) and $\pi^{+}$ (b) multiplicities
of forward and backward rapidities in Au+Au collisions with (labeled as With) and without (labeled as W/O) the consideration of
the induced electric field at a beam energy of 400MeV/nucleon. } \label{pifb1}
\end{figure}
\begin{figure}
\centerline{\includegraphics[width=1.0\columnwidth]{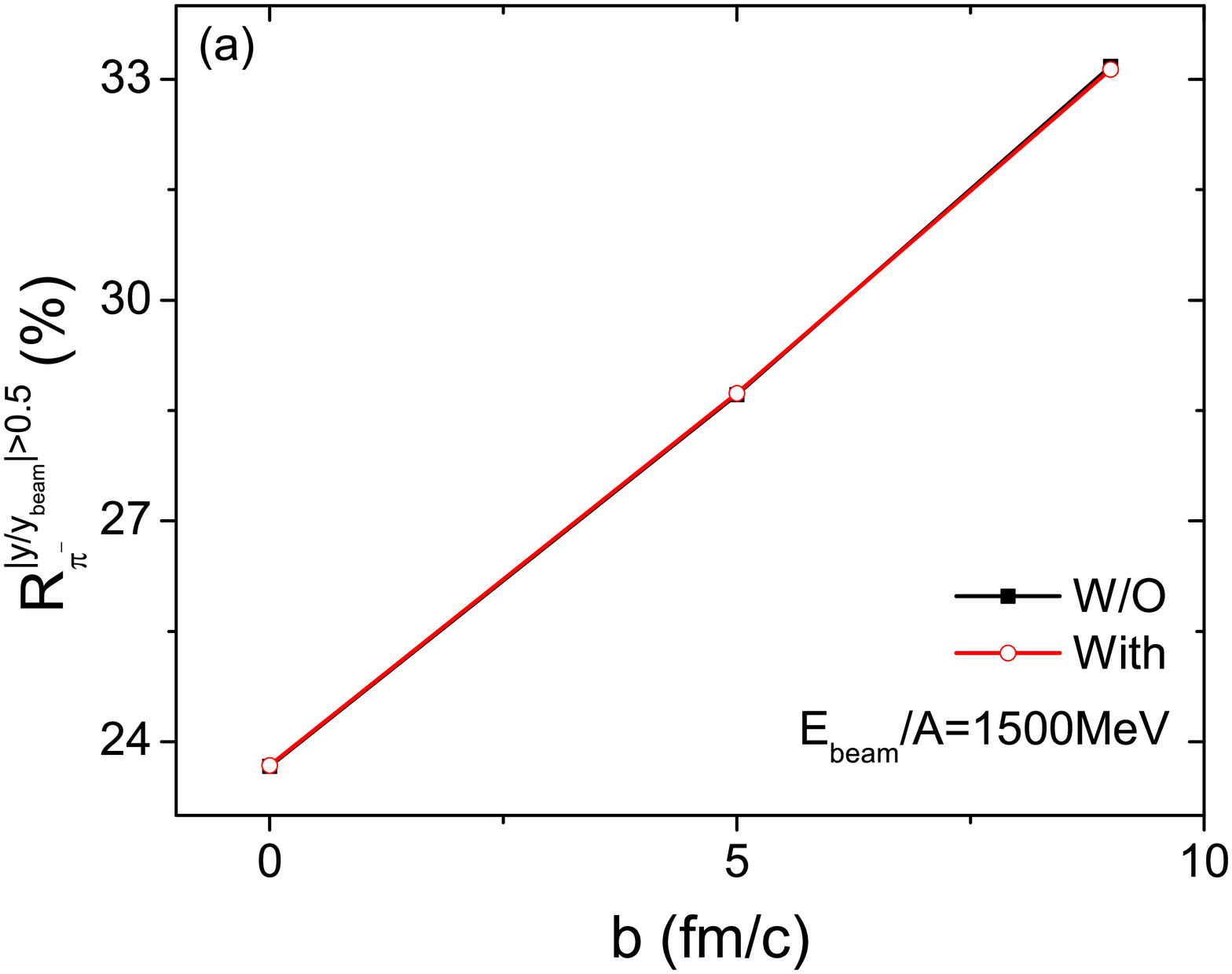}}
\centerline{\includegraphics[width=1.0\columnwidth]{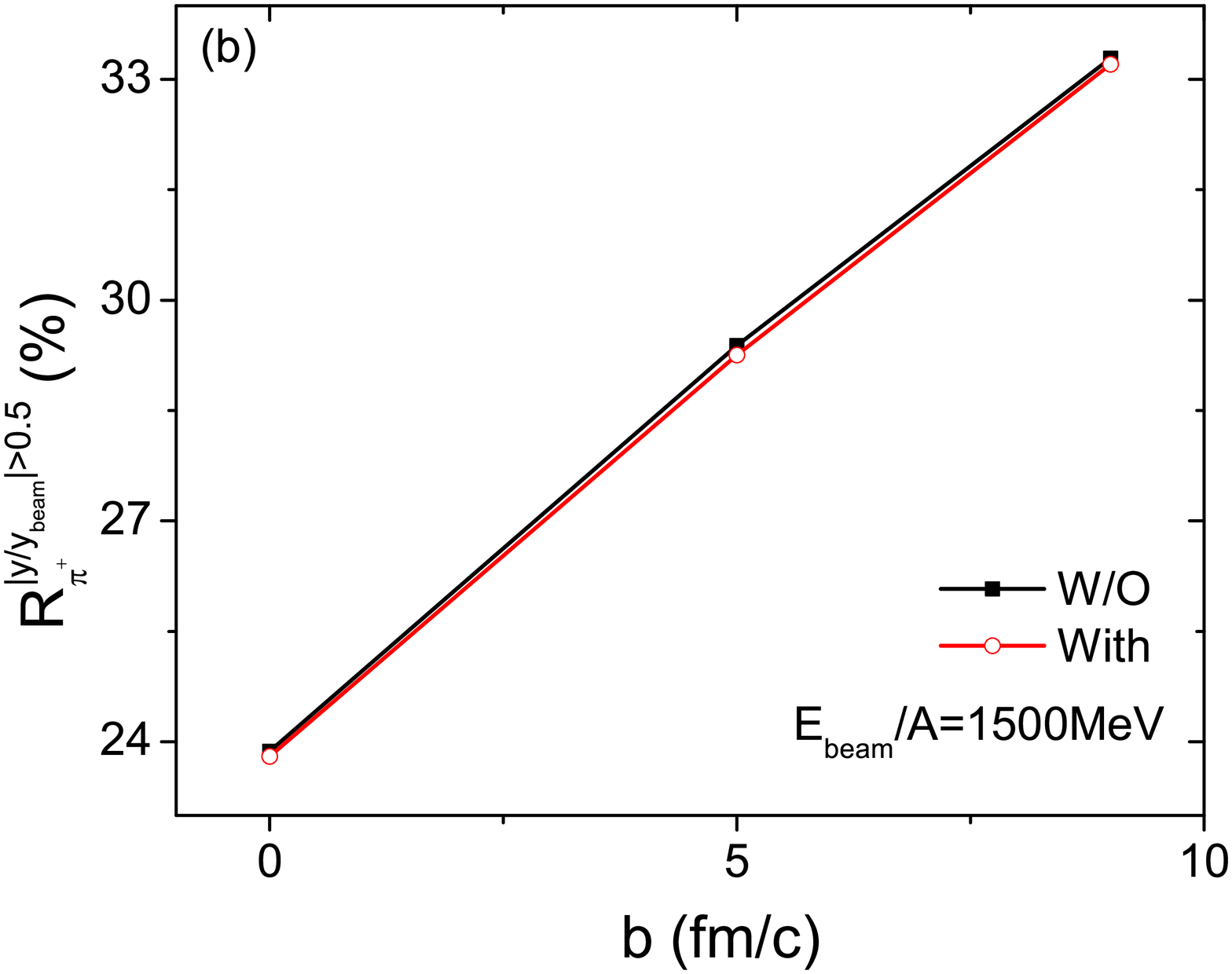}}
 \caption{(Color online) Same as Fig. \ref{pifb1} but for a beam energy of 1500MeV/nucleon. } \label{pifb2}
\end{figure}
Now let's check the impact-parameter-dependent \rpi ratio in Au+Au collisions with and
without the consideration of the induced electric field. Within the IBUU transport model for
heavy-ion collision at the intermediate energy, almost all the pions are produced from
the decay of $\Delta$(1232) resonances. Therefore, the dynamic pion ratio, i.e.,
(\rpi)$_{\rm like}$, can be defined as
\begin{equation}\label{ratio}
(\pi^{-}/\pi^{+})_{\rm like}\equiv
\frac{\pi^{-}+\Delta^{-}+\frac{1}{3}\Delta^{0}}
{\pi^{+}+\Delta^{++}+\frac{1}{3}\Delta^{+}}.
\end{equation}
Because all the $\Delta$ resonances will eventually decay at the final reaction stage, 
the (\rpi)$_{\rm like}$ ratio will naturally become the \rpi ratio. Shown in Fig. \ref{tpion}
is the time evolution of dynamic (\rpi)$_{\rm like}$ ratio from central to peripheral Au+Au collisions
at two beam energies of 400 MeV/nucleon (left panels) and 1500MeV/nucleon (right panels). It can be seen 
that the induced electric field does not affect the total \rpi ratio at either the lower beam energy of 
400MeV/nucleon or the higher beam energy of 1500MeV/nucleon. In fact, because the magnetic field itself 
does not change the total multiplicities of both positive and negative pion mesons as shown in Ref. \cite{Ou11}, 
it therefore can be understood that its derived product, i.e., the induced electric field, is unlikely to 
alter the total \rpi ratio in intermediate-energy heavy-ion collisions . However, just as pointed out in Ref. \cite{Ou11},
the magnetic field has a strong focusing (diverging) effects on positive (negative) pion mesons
at forward (backward) rapidity; therefore, the natural question is whether the induced electric field does
affect the distribution of charged pion mesons and thus the distribution of the \rpi ratio. To this end, we
define the percentage of $\pi^{-}$ and $\pi^{+}$ multiplicities in different rapidities with and without
the consideration of the induced electric field as follows:
\begin{equation}\label{pectge}
R^{|y/y_{\rm beam}|}_{i}\equiv \frac{N^{|y/y_{\rm beam}|}_{i}}{N_{i}}\times \%,
\end{equation}
where index $i$ denotes $\pi^{-}$ or $\pi^{+}$, $y/y_{\rm beam}$ is the reduced rapidity of particle $i$,
$N^{|y/y_{beam}|}_{i}$ denotes the multiplicities of particle $i$ in rapidity interval of
[$-y/y_{\rm beam}$,$y/y_{\rm beam}$], and $N_{i}$ denotes the multiplicities of particle $i$ in all rapidity
range. Shown in Figs. \ref{pimid1} and \ref{pimid2} are the impact-parameter-dependent percentage of
$\pi^{-}$ and $\pi^{+}$ multiplicities in midrapidity ($|y/y_{\rm beam}|\leq 0.5$) at beam energies
of 400 and 1500 MeV/nucleon. It can be seen that the induced electric field obviously reduces
the midrapidity emission of $\pi^{-}$ mesons but enhances the midrapidity emission of $\pi^{+}$ mesons
due to its opposite effects on $\pi^{-}$ and $\pi^{+}$ mesons, especially for the more peripheral and
lower-energy collisions. Within the beam-energy and impact-parameter dependence of the magnetic field as
shown in Fig. \ref{em}, one naturally expects that the effects of the induced electric field on charged
pion mesons are increasing with increasing both the impact parameter and the beam energy. However,
compared to the 400 MeV/nucleon case, the effects of the induced electric field on charged pion mesons
in midrapidity at 1500MeV/nucleon are almost negligible. This is because while the increased
magnetic field with the beam energy may generate a more strong induced electric field, the obviously
decreased duration time of the reaction stage with the beam energy causes the acting time of the induced electric
field on charged pion mesons to become shorter and thus leads to a negligible effect at the beam
energy of 1500MeV/nucleon. Shown in Figs. \ref{pifb1} and \ref{pifb2} are the corresponding impact-parameter-dependent 
percentage of $\pi^{-}$ and $\pi^{+}$ multiplicities in forward and backward rapidities
($|y/y_{\rm beam}|\geq 0.5$) at beam energies of 400 and 1500 MeV/nucleon. Similar to the above
observations from Figs. \ref{pimid1} and \ref{pimid2}, the effects of the induced electric field on charged
pion mesons in forward and backward rapidities at 1500 MeV/nucleon are almost negligible, while for
charged pion mesons in forward and backward rapidities at 400 MeV/nucleon, the induced electric fields
obviously enhance the emission of $\pi^{-}$ meson but reduce the emission of $\pi^{+}$ meson. Naturally,
the total multiplicities of both $\pi^{-}$ and $\pi^{+}$ mesons in all rapidity range and thus their
total \rpi ratio are not affected by the induced electric field as shown in Fig. \ref{tpion}. This feature
indicates that the total \rpi ratio is still a precisely reliable probe of the symmetry energy at both the
lower and higher beam energies. However, for the differential \rpi ratio, due to the fact that the effects of the induced
electric fields on it are obvious at the lower beam energy of 400 MeV/nucleon but negligible at the
higher beam energy of 1500 MeV/nucleon, one naturally needs to look at the effects of the induced electric
fields on their rapidity distribution at the beam energy of 400 MeV/nucleon in the following.

\begin{figure}[h]
\centerline{\includegraphics[width=1.0\columnwidth]{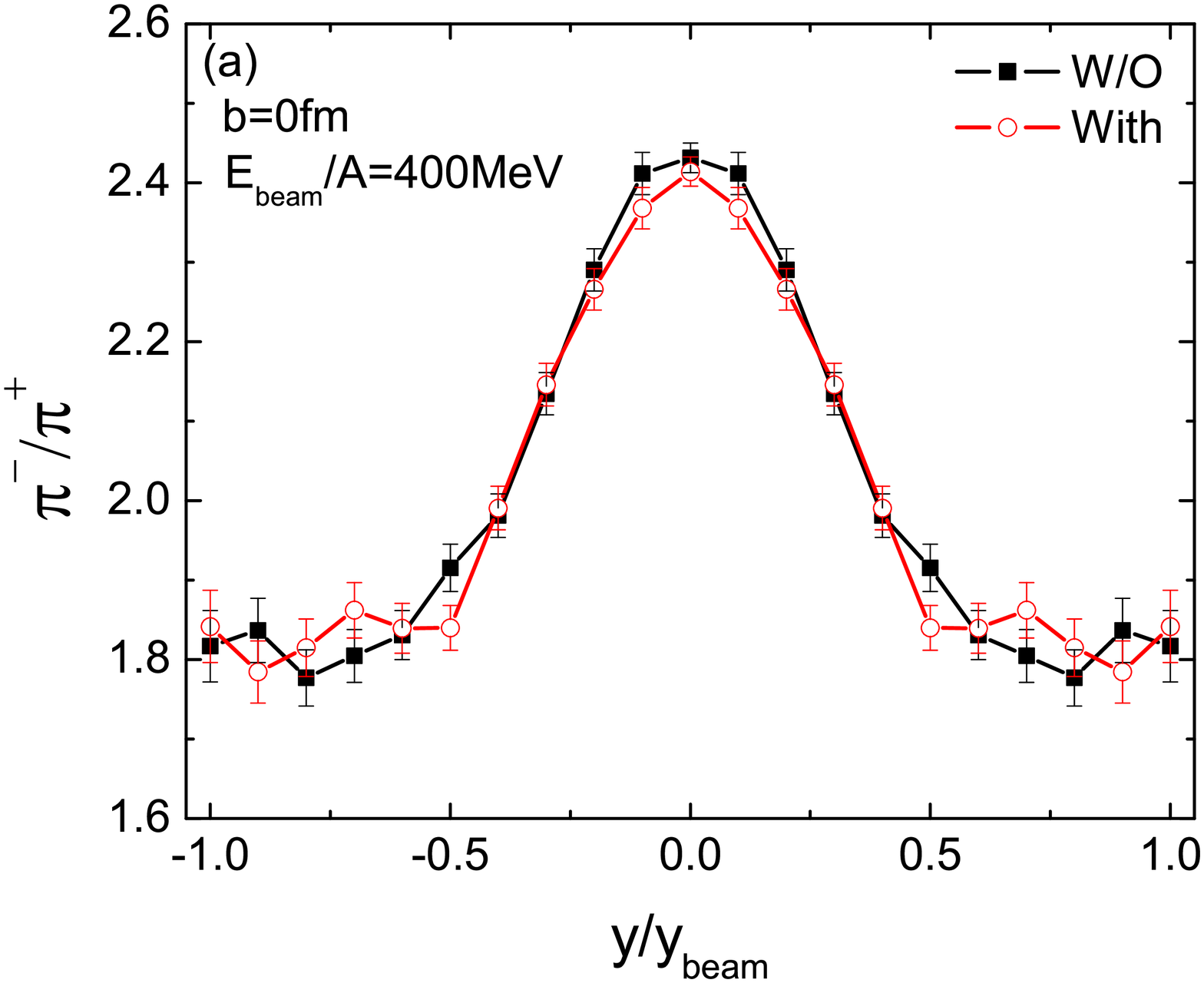}}
\centerline{\includegraphics[width=1.0\columnwidth]{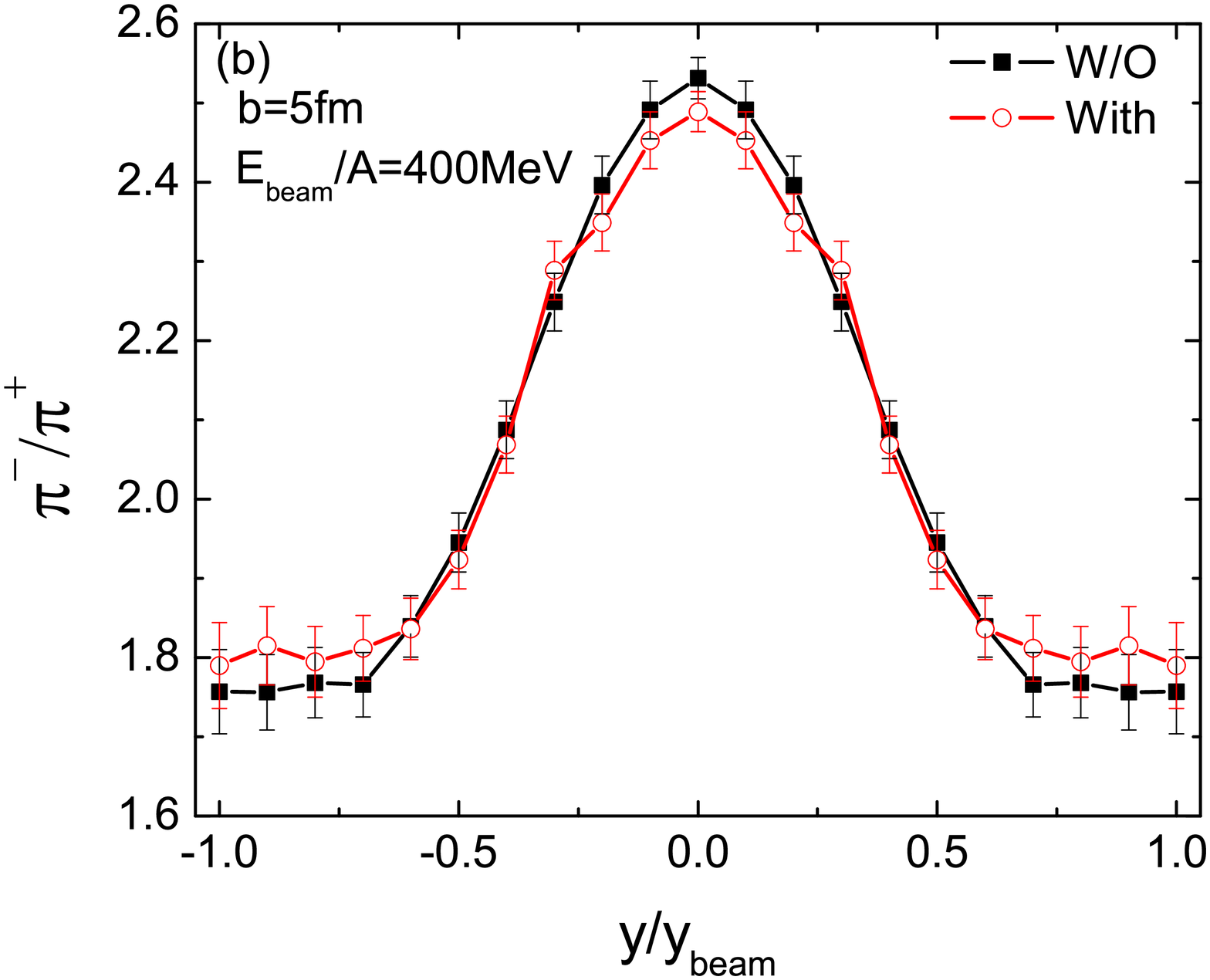}}
\centerline{\includegraphics[width=1.0\columnwidth]{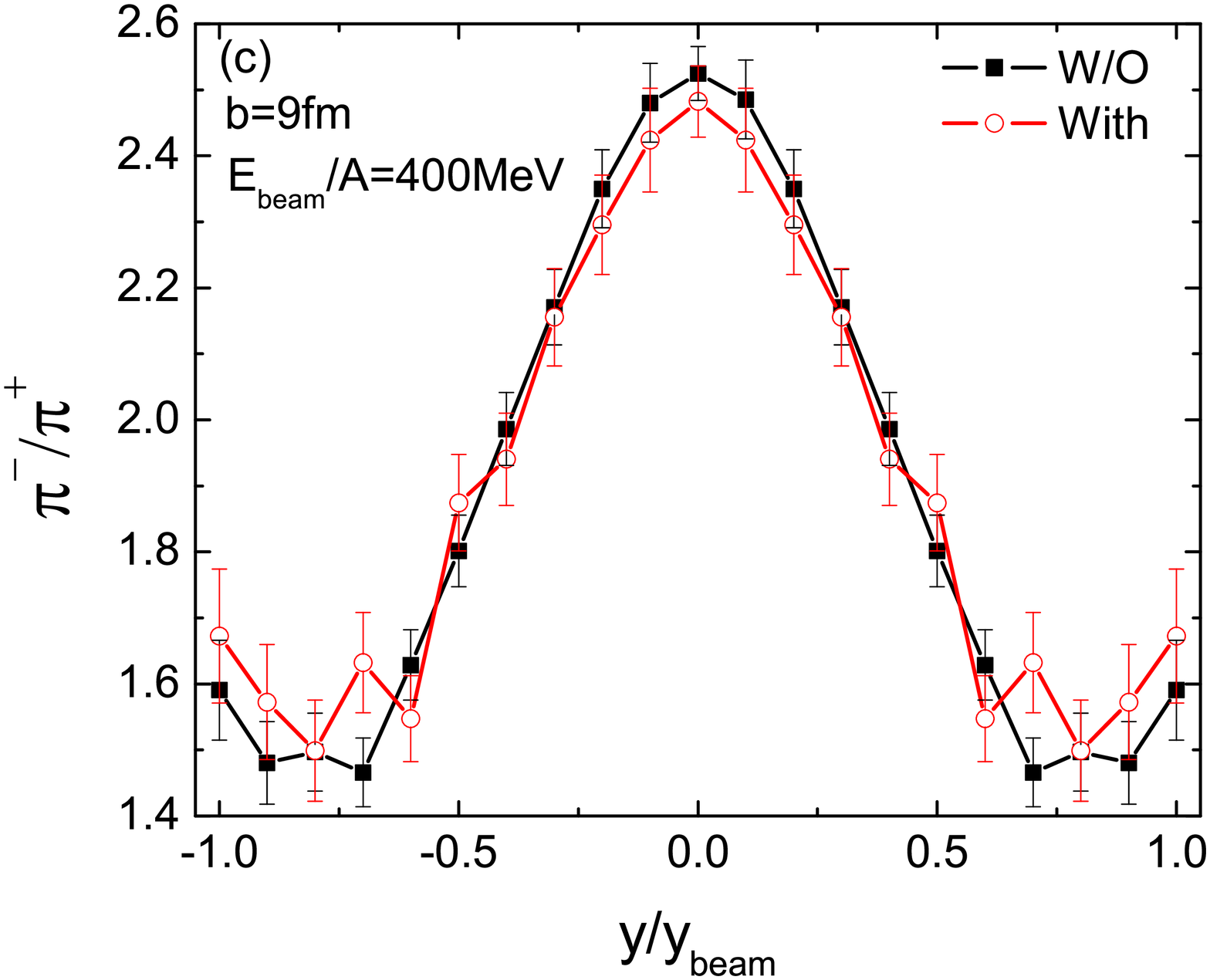}}
\caption{(Color online) The rapidity distribution of the \rpi ratio in
central to peripheral Au+Au collisions with (labeled as With) and without (labeled as W/O) the consideration of
the induced electric field at a beam energy of 400 MeV/nucleon.}\label{rapidity}
\end{figure}

\begin{figure}[h]
\centerline{\includegraphics[width=1.2\columnwidth]{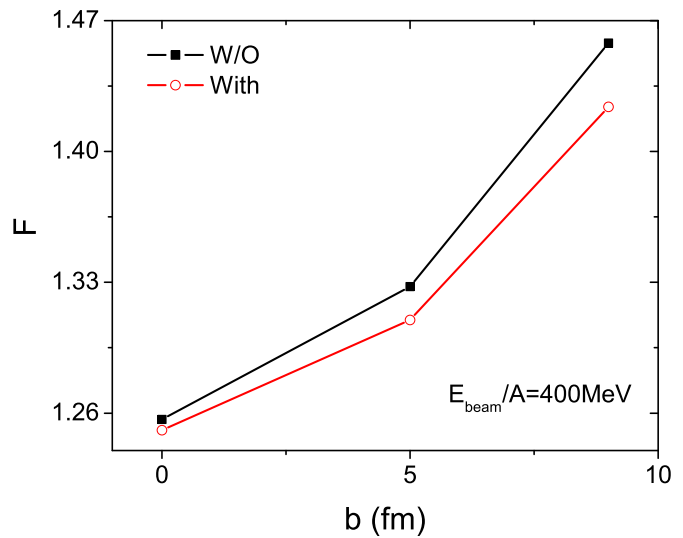}}
\caption{(Color online) The impact-parameter-dependent relative suppression factor $F$
in Au+Au collision with (labeled as With) and without (labeled as W/O) the consideration of
the induced electric field at a beam energy of 400 MeV/nucleon. } \label{F}
\end{figure}

Shown in Fig. \ref{rapidity} is the rapidity distribution of the \rpi ratio in central to peripheral
Au+Au collisions with and without the consideration of the induced electric field at the beam energy of
400 MeV/nucleon. It can be found that the induced electric fields indeed suppress the \rpi ratio in
midrapidity but enhance the \rpi ratio in forward and backward rapidities especially for more
peripheral collisions. This feature indicates that the differential \rpi ratio is less delicate
than the total \rpi ratio in probing the symmetry energy. Finally, to show more explicitly the
effects of the induced electric fields on the differential \rpi ratio, we define a relative suppression
factor as follows:
\begin{equation}
F\equiv \frac{(\pi^{-}/\pi^{+})_{|y/y_{\rm beam}|\leq 0.5}}{(\pi^{-}/\pi^{+})_{|y/y_{\rm beam}|\geq 0.5}}.
\end{equation}
Obviously, $F$ will more clearly indicate the effects of the induced electric fields on the differential
\rpi ratio because this ratio method will maximize the effects of the induced electric fields
on the \rpi ratio due to its suppression effects on the \rpi ratio in midrapidity but enhancing effects on
the \rpi ratio in forward and backward rapidity. On the other hand, considering that this ratio method can
minimize the effects from both the in-medium nucleon-nucleon cross section and the poorly known symmetry
energy due to the pion mesons in different rapidities approximately experiencing the same influences from
them, $F$ thus can be an effective probe of the induced electric fields generated in heavy-ion
collisions. Show in Fig. \ref{F} is the relative suppression factor in central to peripheral Au+Au
collisions at a beam energy of 400MeV/nucleon. It can be seen that the relative suppression factor
$F$ increases with the impact parameter increasing due to the steeper magnetic field generating
a stronger induced electric field in peripheral collisions. This feature indicates that the relative
suppression factor $F$ can indeed be an effective probe of induced electric fields generated
in heavy-ion collisions.

\section{Summary}\label{summary}
In summary, we have carried out an investigation about the effects of the electric field due to
a variable magnetic field on the \rpi ratio in central to peripheral heavy-ion collisions. Within an
isospin- and momentum-dependent transport model, the Au+Au collisions are performed at two beam
energies of 400 and 1500 MeV/nucleon. It is shown that, while the induced electric
field does not affect the total \rpi ratio at both the lower and higher beam energies, it
suppresses the differential \rpi ratio in midrapidity but enhances the differential \rpi ratio
in forward and backward rapidities due to the reduced (enhanced) emission of $\pi^{-}$ ($\pi^{+}$)
meson in midrapidity but the enhanced (reduced) emission of $\pi^{-}$ ($\pi^{+}$) meson in forward
and backward rapidities. These findings imply that the total \rpi ratio is still a precisely
reliable probe of symmetry energy, but the induced electric field should be considered in
future studies of using the differential \rpi ratio as the probe of symmetry energy. Moreover,
because the relative suppression factor can not only minimize the effects of some uncertainty factors in
heavy-ion collisions but also can maximize the effects of the induced electric field, therefore we may conclude
that the relative suppression factor can be an effective probe of the induced electric field
generated in heavy-ion collisions in future studies.\\

\noindent{\textbf{Acknowledgements}} \\
G.F. Wei is grateful to Prof. Bao-An Li for stimulating the work on this project and Prof.
Li Ou for helpful discussions. The author is also grateful for the help provided by the High-Performance
Computational Science Research Cluster at Texas A$\&$M University-Commerce where partial
calculations of this work were done. This work is supported by the National Natural Science
Foundation of China under grant No.11405128 and in part by 20160978-SIP-IPN, Mexico.

\end{CJK*}


\begin{thebibliography}{99}
\bibitem{FOPI} W. Reisdorf {\it et al.}, Nucl. Phys. A \textbf{781}, 459 {2007}; Nucl. Phys. A 848, 366 (2010).
\bibitem{FOPI-LAND1} Y. Leifels {\it et al.}, Phys.Rev.Lett. 71, 963 (1993).
\bibitem{FOPI-LAND2} D. Lambrecht {\it et al.}, Z.Phys. A350, 115 (1994).
\bibitem{Xiao09} Z. G. Xiao, B. A. Li, L. W. Chen, G. C. Yong, and M. Zhang, Phys. Rev. Lett. \textbf{102}, 062502 (2009).
\bibitem{Xie13} W. J. Xie, J. Su, L. Zhu, and F. S. Zhang, Phys. Lett. B \textbf{718}, 1510 (2013).
\bibitem{Feng10} Z. Q. Feng, G. M. Jin, Phys. Lett. B \textbf{683}, 140 (2010).
\bibitem{Hong14}J. Hong, P. Danielewicz, Phys. Rev. C \textbf{90}, 024605 (2014).
\bibitem{Guo15a}W. M. Guo, G. C. Yong, H. Liu, and W. Zuo, Phys. Rev. C \textbf{91}, 054616 (2015).
\bibitem{Bao15a}B. A. Li, Phys. Rev. C \textbf{92}, 034603 (2015).
\bibitem{Guo15b}W. M. Guo, G. C. Yong, W. Zuo, Phys. Rev. C \textbf{92}, 054619 (2015).
\bibitem{Song15}T. Song, C. M. Ko, Phys. Rev. C \textbf{91}, 014901 (2015).
\bibitem{Cozma16}M. D. Cozma, Phys. Lett. B \textbf{753}, 166 (2016).
\bibitem{Orhen15}O. Hen, B. A. Li, W. J. Guo, L. B. Weinstein, E. Piasetzky, Phys. Rev. C \textbf{91}, 025803 (2015).
\bibitem{Bao15b}B. A. Li, W. J. Guo, Z. Z. Shi, Phys. Rev. C \textbf{91}, 044601 (2015).
\bibitem{Wei14}G. F. Wei, B. A. Li, J. Xu, and L. W. Chen, Phys. Rev. C \textbf{90}, 014610 (2014).
\bibitem{Wang13}G.Wang {\it et al.}, Nucl. Phys. A 904-905, 248c (2013).
\bibitem{Ke12}H. Ke {\it et al.}, J. Phys. Conf. Ser. \textbf{389}, 012035 (2012).
\bibitem{Adam15}L. Adamczyk {\it et al.}, Phys. Rev. Lett. \textbf{114}, 252302 (2015).
\bibitem{Huang15}X. G. Huang, arXiv:1509.04073.
\bibitem{Ou11}L. Ou, B. A. Li, Phys. Rev. C \textbf{84}, 064605 (2011).
\bibitem{Deng12}W. T. Deng, X. G. Huang, Phys. Rev. C \textbf{85}, 044907 (2012).
\bibitem{Liu12}K. F. Liu, Phys. Rev. C \textbf{85}, 014909 (2012).
\bibitem{Bar05}V. Baran, M. Colonna, V. Greco, and M. Di toro, Phys. Rep. \textbf{410}, 335 (2005).
\bibitem{Bao08}B. A. Li, L. W. Chen, C. M. Ko, Phys. Rep. \textbf{464}, 113 (2008).
\bibitem{IBUU}B. A. Li, C. B. Das, S. Das Gupta, and C. Gale, Phys. Rev. C \textbf{69} (2004) 011603(R).
\bibitem{Das03} C. B. Das, S. Das Gupta, C. Gale, and B. A. Li, Phys. Rev. C \textbf{67}, 034611 (2003).
\bibitem{Che05} L. W. Chen, C. M. Ko, and B. A. Li, Phys. Rev. Lett. \textbf{94}, 032701 (2005).


\end{thebibliography}
\end{document}